\documentclass[a4paper,11pt]{article}
\usepackage{amsmath}
\usepackage{epsfig}
\usepackage[round]{natbib}

\title{Getting Your Eye In: A Bayesian Analysis of Early Dismissals in Cricket}
\author{Brendon James Brewer\\
        School of Mathematics and Statistics\\
        The University of New South Wales\\
        \\
        \texttt{brendon.brewer@unsw.edu.au}}
\date{\today}

\begin{document}

\maketitle

\abstract{A Bayesian Survival Analysis method is motivated and developed for analysing sequences of scores made by a batsman in test or first class cricket. In particular, we expect the presence of an effect whereby the distribution of scores has more probability near zero than a geometric distribution, due to the fact that batting is more difficult when the batsman is new at the crease. A Metropolis-Hastings algorithm is found to be efficient at estimating the proposed parameters, allowing us to quantify exactly how large this early-innings effect is, and how long a batsman needs to be at the crease in order to ``get their eye in''. Applying this model to several modern players shows that a batsman is typically only playing at about half of their potential ability when they first arrive at the crease, and gets their eye in surprisingly quickly. Additionally, some players are more ``robust'' (have a smaller early-innings effect) than others, which may have implications for selection policy.}

\section{Introduction}
It is well known to cricketers of all skill levels that the longer a batsman is in for, the easier batting tends to become. This is probably due to a large number of psychological and technique-related effects: for example, it is generally agreed that it takes a while for a batsman's footwork to ``warm up'' and for them to adapt to the subtleties of the prevailing conditions and the bowling attack. Consequently, it is frequently observed that players are far more likely to be dismissed early in their innings than is predicted by a constant-hazard model, where the probability of getting out on your current score (called the {\it Hazard}) is exactly the same regardless of your current score. Note that a constant hazard model leads to an exponential probability distribution over the non-negative integers (i.e. the geometric distribution) as describing our prediction of a batsman's score. The aim of this paper is to develop a Bayesian method \citep{2004kats.book.....O} for inferring how a player's Hazard varies throughout an innings, thus giving quantitative answers to the questions ``how long do we have to wait until batsmen get their eye in, and how much better do they really become?''. This question has been addressed previously using nonparametric frequentist survival analysis \citep{frequentist,cai}. However, using a nonparametric approach in a Bayesian setting would give the hazard function far too much freedom and would lead to very poorly constrained inferences of the hazard function if applied to individual players. To simplify matters, this paper uses a parametric model, which is effectively a single change-point model with a smooth transition rather than a sudden jump.

\section{Sampling Distribution}
Consider predicting the score $X \in \{0,1,2,3,...\}$ that a batsman will make in a single innings. We will now assign a probability distribution for $X$, conditional on some parameters. Define a hazard function $H(x) \in [0,1]$ as the probability of being dismissed on score $x$ (i.e. $P(X=x)$) given that the batsman is currently on score $x$ (i.e. given $X \geq x$):
\begin{equation}\label{hazardDefinition}
H(x) = P(X = x | X \geq x) = \frac{P(X = x, X\geq x)}{P(X \geq x)} = \frac{P(X = x)}{P(X \geq x)}
\end{equation}
Define a backwards cumulative distribution function by:
\begin{equation}
G(x) = P(X \geq x)
\end{equation}
Using $G(x)$ rather than the conventional cumulative distribution $F(x)=P(X \leq x)$ simplifies some expressions in this case, and also helps because $G(x)$ will also serve as the likelihood function for the ``not-outs'', or uncompleted innings. With this definition, Equation~\ref{hazardDefinition} becomes, after some rearrangement, a difference equation for $G$:
\begin{equation}
G(x+1) = \left[1 - H(x)\right]G(x)
\end{equation}
With the initial condition $G(0) = 1$, this can be solved, giving:
\begin{equation}
G(x) = \prod _{a=0}^{x-1}\left[1 - H(a)\right]
\end{equation}
This is the product of the probabilities of surviving to score $1$ run, times the probability of reaching a score of $2$ runs given that you scored $1$, etc, up to the probability of surviving to score $x$ given that you scored $x-1$. Thus, the probability distribution for $X$ is given by the probability of surviving up to a score $x$ and then being dismissed:
\begin{equation}
P(X = x) = H(x)\prod _{a=0}^{x-1}\left[1 - H(a)\right]
\end{equation}
This is all conditional on a choice of the hazard function $H$, which we will parameterise by parameters $\theta$. Assuming independence (this is not a physical assertion, rather, an acknowledgement that we are not interested in any time dependent effects for now), the probability distribution for a set of scores $\{x_i\}_{i=1}^{I-N}$ ($I$ and $N$ are the number of innings and not-outs respectively) and a set of not-out scores $\{y_i\}_{i=1}^N$ is:
\begin{equation}\label{likelihood}
p(\mathbf{x},\mathbf{y}|\theta) = \prod_{i=1}^{I-N}\left(H(x_i;\theta)\prod _{a=0}^{x_i-1}\left[1 - H(a;\theta)\right]\right) \times \prod_{i=1}^{N}\left(\prod _{a=0}^{y_i-1}\left[1 - H(a;\theta)\right]\right)
\end{equation}
When the data $\{\mathbf{x},\mathbf{y}\}$ are fixed and known, Equation~\ref{likelihood} gives the likelihood for any proposed model of the Hazard function - that is, for any value of $\theta$. The log likelihood is:
\begin{equation}
\log p(\mathbf{x},\mathbf{y}|\theta) = \sum_{i=1}^{I-N} \log H(x_i;\theta) + \sum_{i=1}^{I-N} \sum_{a = 0}^{x_i - 1} \log \left[1 - H(a;\theta) \right] + \sum_{i=1}^{N} \sum_{a = 0}^{y_i - 1} \log \left[1 - H(a;\theta) \right]
\end{equation}

\section{Parameterisation of the Hazard Function}
Rather than seek clever parameterisations of $H(x;\theta)$ and priors over $\theta$ that are conjugate to the likelihood, we will take the simpler approach of simply defining a model and prior, and doing the inference with a Metropolis-Hastings sampler (C++ source code and data files for carrying this out will be provided by the author on request). To capture the phenomenon of ``getting your eye in'', the Hazard function will need to be high for low $x$ and decrease to a constant value as $x$ increases and the batsman becomes more comfortable. Note that if $H(x) = h$, a constant value, the sampling distribution becomes a geometric distribution with expectation $\mu = 1/h - 1$. This suggests modelling the Hazard function in terms of an ``effective batting average'' that varies with time, which is helpful because it is easier to think of playing ability in terms of batting averages than dismissal probabilities. $H(x)$ is obtained from $\mu(x)$ as follows:
\begin{equation}
H(x) = \frac{1}{\mu(x) + 1}
\end{equation}
A simple change-point model for $\mu(x)$ would be to have $\mu(x) = \mu_1 + (\mu_2 - \mu_1)\textnormal{Heaviside}(x - \tau)$, where $\tau$ is the change-point. However, a more realistic model would have $\mu$ changing smoothly from one value to the other. Replacing the Heaviside step function with a logistic sigmoid function of the form $1/(1 + e^{-t})$ gives the following model, which will be adopted throughout this paper:
\begin{equation}\label{model}
\mu(x) = \mu_1 + \frac{\mu_2 - \mu_1}{1 + \exp \left(-(x-\tau) / L \right)}
\end{equation}
Hence
\begin{equation}
H(x) = \left[1 + \mu_1 + \frac{\mu_2 - \mu_1}{1 + \exp \left(-(x-\tau) / L \right)}\right]^{-1}
\end{equation}
This has four parameters: $\mu_1$ and $\mu_2$, the two effective abilities of the player, $\tau$, the midpoint of the transition between them, and $L$, which describes how abrupt the transition is. As $L \to 0$ this model resembles the simpler change-point model. A few examples of the kind of hazard models that can be described by varying these parameters are shown in Figure~\ref{models}. It is possible (although we don't really expect it) for the risk of being dismissed to increase as your score increases; more commonly it will decrease. Slow or abrupt transitions are possible and correspond to different values of $L$.

\section{Prior Distribution}\label{priors}
\begin{figure}
\begin{center}
\includegraphics[scale=1]{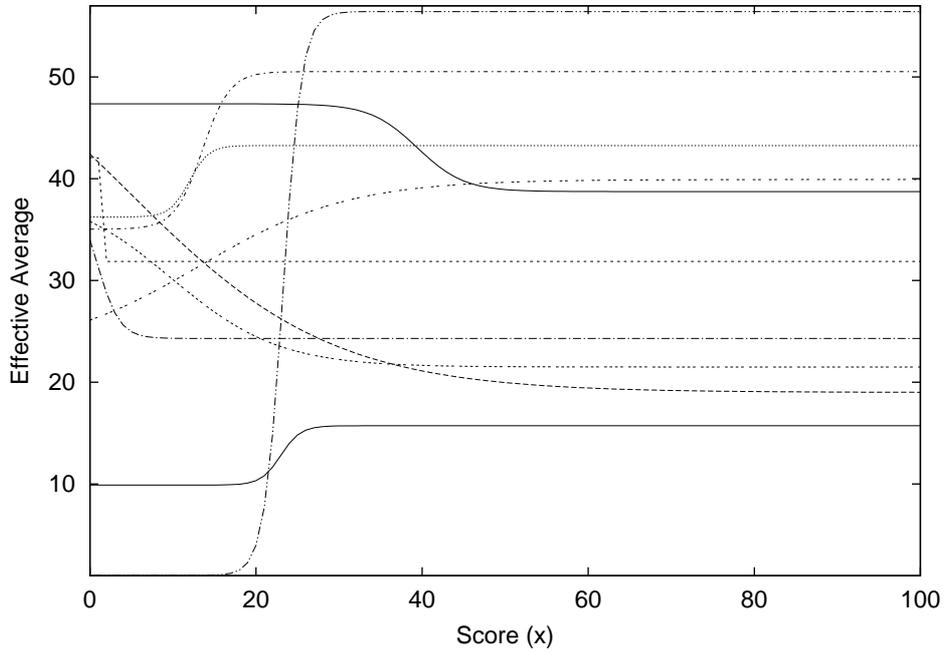}
\end{center}
\caption{Illustrative examples of the kind of functions produced by Equation~\ref{model}, with a range of typical values (chosen from the prior, see Section~\ref{priors}) of the four parameters $\mu_1$, $\mu_2$, $\tau$ and $L$. \label{models}}
\end{figure}

Now we must assign a prior probability distribution of the space of possible values for the parameters $(\mu_1, \mu_2, \tau, L)$. All of these parameters are non-negative and can take any positive real value. It is possible to take into account prior correlations between $\mu_1$ and $\mu_2$ (describing the expection that a player who is excellent when set is also more likely to be good just after arriving at the crease, and that $\mu_2$ is probably greater than $\mu_1$). However, this will almost certainly be supported by the data anyway. Hence, for simplicity we assigned the more conservative independent $\textnormal{Normal}(30,20^2)$ priors\footnote{A reasonable first-order description of the range of expected variation in batting abilities and hence our state of knowledge about a player whose identity is unspecified - we intend the algorithm to apply to any player. It is possible to parameterise this prior with unknown hyperparameters and infer them from the career data of many players, yielding information about the cricket population as a whole. However, such a calculation is beyond the scope of this paper.}, truncated to disallow negative values:
\begin{equation}\label{prior1}
p(\mu_1,\mu_2) \propto \exp \left[-\frac{1}{2}\left(\frac{\mu_1 - 30}{20}\right)^2 -\frac{1}{2}\left(\frac{\mu_2 - 30}{20}\right)^2\right] \hspace{2cm} \mu_1,\, \mu_2 > 0
\end{equation}
The joint prior for $L$ and $\tau$ is chosen to be independent of the $\mu$'s and also independent of each other. A typical player can expect to become accustomed to the batting conditions after $\sim$ 20 runs. An exponential prior for $\tau$ with mean 20 and an exponential prior with mean 3 to $L$ were found to produce a range of plausible hazard functions:
\begin{equation}\label{prior2}
p(\tau,L) \propto \exp\left(-\frac{\tau}{20}-\frac{L}{3}\right) \hspace{2cm} \tau,\,L > 0
\end{equation}
Some hazard functions sampled from the prior are displayed in Figure~\ref{models}. The posterior distribution for $\mu_1, \mu_2, \tau$ and $L$ is proportional to \texttt{prior} $\times$ \texttt{likelihood}, i.e. the product of the right hand sides of Equations~\ref{likelihood},~\ref{prior1} and~\ref{prior2}. Qualitatively, the effect of Bayes' theorem is to take the set of possible hazard functions and their probabilities (Figure~\ref{models}) and reweight the probabilities according to how well each hazard function fits the observed data.

\section{The Data}
Data were obtained from the StatsGuru utility on the Cricinfo website (\texttt{http://www.cricinfo.com/}) for the following players: Brian Lara, Chris Cairns, Nasser Hussain, Gary Kirsten, Justin Langer, Shaun Pollock, Steve Waugh and Shane Warne. These players were chosen arbitrarily but subject to the condition of having recently completed long careers. A selection of batsmen, quality all-rounders and bowlers was chosen. The MCMC was run for a large number of steps - mixing is quite rapid because the likelihood evaluation is fast and the parameter space is only 4-dimensional. For brevity, we will display posterior distributions for Brian Lara only. For the other players, summaries such as the posterior means and standard deviations will be displayed instead.

\section{Results}
\subsection{Marginal Posterior Distributions}
\begin{figure}
\begin{center}
\includegraphics[scale=0.5]{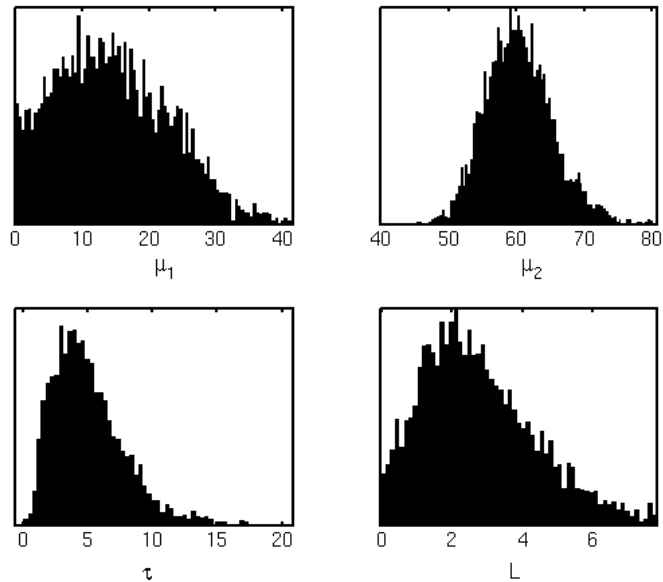}
\end{center}
\caption{Results for the four parameters for Brian Lara. The top two panels show the posterior distributions for his two abilities (effective batting averages) $\mu_1$ and $\mu_2$, while the lower panels show the distributions for the change-point $\tau$ and change-timescale $L$. See the text for interpretation. \label{lara_results}}
\end{figure}

In this section we will focus on the posterior distributions of the parameters $(\mu_1, \mu_2, \tau, L)$ for Brian Lara. The marginal distributions (approximated by samples from the MCMC simulation) are plotted in Figure~\ref{lara_results}. These results imply that when Lara is new to the crease, he bats like a player with an average of $\sim$ 15, until he has scored $\sim$ 5 runs. After a transition period with a scale length of $\sim$ 2 runs, (although the form of the logistic functions shows that a transition is more gradual than indicated by $L$), he then begins to bat as though he has an average of $\sim$ 60. In this case, the analysis has confirmed the folklore about Brian Lara - that if you don't get him out early, you can never really tell when he might get a huge score. ``Form'' doesn't really come into it. The only surprise to emerge from this analysis is the low value of the change-point $\tau$ - Lara is halfway through the process of getting his eye in after scoring only about 5 runs. However, there is still a reasonable amount of uncertainty about the parameters, even though Brian Lara's long test career consisted of 232 innings. The posterior distribution for Brian Lara's parameters does not contain strong correlations (the maximum absolute value in the correlations matrix is 0.4). This is also true of the posterior distributions for the other players. Hence, in the next section, summaries of the marginal distributions for the four parameters will be presented for each player.

\subsection{Summaries}
\begin{figure}
\begin{center}
\caption{Estimated location of each of the players on the $\mu_1$-$\mu_2$ plane. The line corresponds to $\mu_2=\mu_1$, and the closer a player is to the line, the more robust the player. Surprisingly, Shaun Pollock and Chris Cairns lie closest to the line. Although each player has been represented by a point on this plot, these are only estimates (posterior means), and each point is actually just the centre of a large zone of uncertainty.\label{points}}
\includegraphics[scale=1.2]{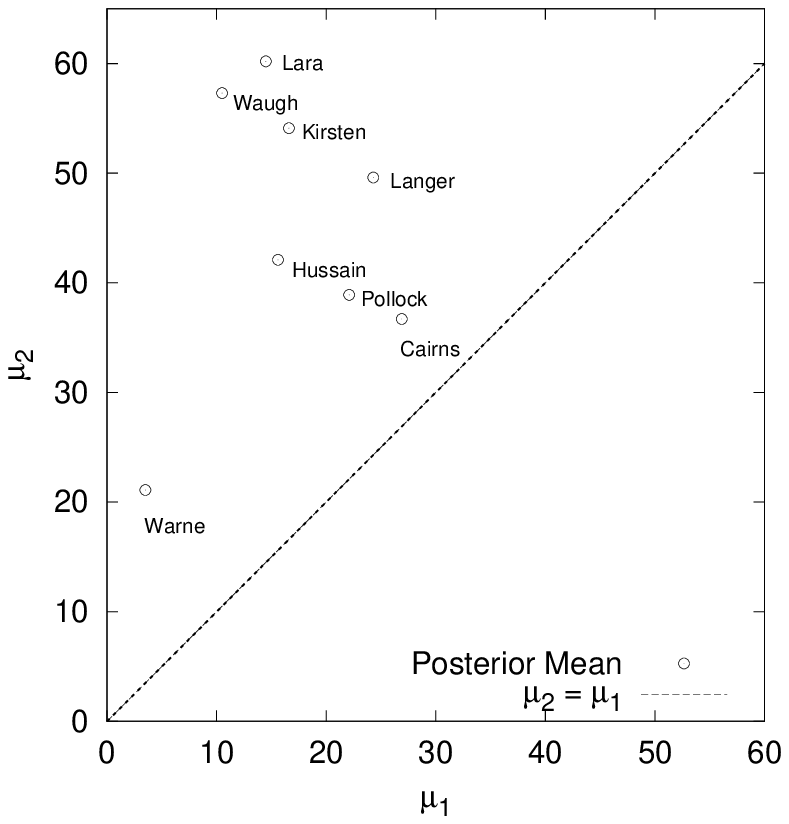}
\end{center}
\end{figure}

The estimates and uncertainties (posterior mean $\pm$ standard deviation) for the four parameters are presented in Table~\ref{summaries}. Figure~\ref{points} also shows graphically where each of the eight players is estimated to lie on the $\mu_1$-$\mu_2$ plane. One interesting result that is evident from this analysis is that it is not just the gritty specialist batsmen that are robust in the sense that $\mu_1$ is quite high compared to $\mu_2$. The two aggressive allrounders Shaun Pollock and Chris Cairns also show this trait, and are even more robust than, for example, Justin Langer and Gary Kirsten. It is possible that the technique or mindset shown by these players is one that does not require much warming up, or that it is more difficult to get your eye in at the top of the order than in the middle/lower order - although this ought to be a very tentative conclusion given that it is based on only two examples.

The estimated value of $\mu_1$ for Steve Waugh is lower than all other players in the sample apart from Shane Warne. Even Shaun Pollock appears to be a better batsman than Steve Waugh at the beginning of his innings. The plausibility of this statement can be measured by asking the question ``what is the posterior probability that $H_{Pollock}(0) < H_{Waugh}(0)$?''. From the MCMC output, this probability was found to be 0.92.

The marginal likelihood or ``evidence'' for this entire model and choice of priors can be measured effectively using annealed importance sampling, or AIS \citep{ais,1997PhRvE..56.5018J,1997PhRvL..78.2690J}. AIS is a very generally applicable MCMC-based algorithm that produces an unbiased estimate of $Z = \int \texttt{prior}(\theta) \times \texttt{likelihood}(\theta) \,d\theta$. $Z$ is the probability of the data that were actually observed, under the model, averaged over all possible parameter values (weighted according to the prior). It is the crucial quantity for updating a state of knowledge about which of two distinct models is correct \citep{mackay,2004kats.book.....O}. To test our model for the hazard function, we computed the evidence value for each player for the varying-hazard model and also for a constant hazard model, with a truncated $N(30,20^2)$ prior on the constant effective average. The logarithm of the Bayes Factor (evidence ratio) describing how well the data support the varying-hazard model over constant hazard is shown in the right-hand column of Table~\ref{summaries}. Since these were computed using a Monte-Carlo procedure, they are not exact, but the AIS simulations were run for long enough so that the standard error in the Bayes Factor for each player was less than 5\% of its value. The data decisively favour the varying-hazard model in all cases, and this would be expected to persist under slight changes to the hazard function parameterisation or the prior distributions.

\begin{table}
\caption{Parameter estimates for the players studied in this paper. The right hand column, the logarithm of the Bayes Factor, shows that the data support the varying hazard model over a constant hazard model by a large factor in all cases. The smallest Bayes Factor is still over 2500 to 1 in favour of the varying Hazard Model. Thus, the varying hazard model would likely still be significantly favoured even if the priors for the parameters were slightly modified.\label{summaries}}\vspace{0.5cm}
\centering
\begin{tabular}{l c c c c | c c}
\hline\hline
Player & $\mu_1$ & $\mu_2$ & $\tau$ & $L$ & log$_e(Z)$ & log$_{e}(Z/Z_0)$\\
\hline
Cairns & 26.9 $\pm$ 9.2 & 36.7 $\pm$ 5.5 & 14.5 $\pm$ 17.7 & 3.1 $\pm$ 3.0 & -444.11 & 8.82\\
Hussain & 15.6 $\pm$ 9.1 & 42.1 $\pm$ 4.4 & 5.2 $\pm$ 7.1 & 2.2 $\pm$ 1.0 & -707.16 & 12.28\\
Kirsten & 16.6 $\pm$ 9.3 & 54.1 $\pm$ 5.7 & 7.3 $\pm$ 5.5 & 2.9 $\pm$ 2.4 & -757.16 & 16.94\\
Langer & 24.3 $\pm$ 11.5 & 49.6 $\pm$ 4.9 & 8.9 $\pm$ 14.3 & 2.8 $\pm$ 2.9 & -810.34 & 11.66\\
Lara & 14.5 $\pm$ 8.3 & 60.2 $\pm$ 4.7 & 5.1 $\pm$ 2.9 & 2.8 $\pm$ 1.8 & -1105.65 & 21.62\\
Pollock & 22.1 $\pm$ 7.7 & 38.9 $\pm$ 5.4 & 9.7 $\pm$ 9.3 & 3.1 $\pm$ 2.9 & -519.39 & 7.87\\
Warne & 3.5 $\pm$ 2.0 & 21.1 $\pm$ 2.0 & 1.1 $\pm$ 0.6 & 0.5 $\pm$ 0.4 & -686.59 & 22.54\\
Waugh & 10.5 $\pm$ 5.5 & 57.3 $\pm$ 4.4 & 1.8 $\pm$ 1.6 & 0.8 $\pm$ 1.2 & -1030.69 & 25.29\\
\hline
{\bf Prior} & 32.8 $\pm$ 17.6 & 32.8 $\pm$ 17.6 & 20.0 $\pm$ 20.0 & 3.0 $\pm$ 3.0 & N/A & N/A
\end{tabular}
\end{table}

\subsection{Predictive Hazard Function}
In the usual way \citep{2004kats.book.....O}, a predictive distribution for the next data point (score in the next innings) can be found by averaging the sampling distribution (Equation~\ref{likelihood}) over all possible values of the parameters that are allowed by the posterior. Of course, all of these players have retired, so this prediction is simply a conceptual device to get a single distribution over scores, and hence a single estimated hazard function via Equation~\ref{hazardDefinition}.

\begin{figure}
\begin{center}
\includegraphics[scale=0.5]{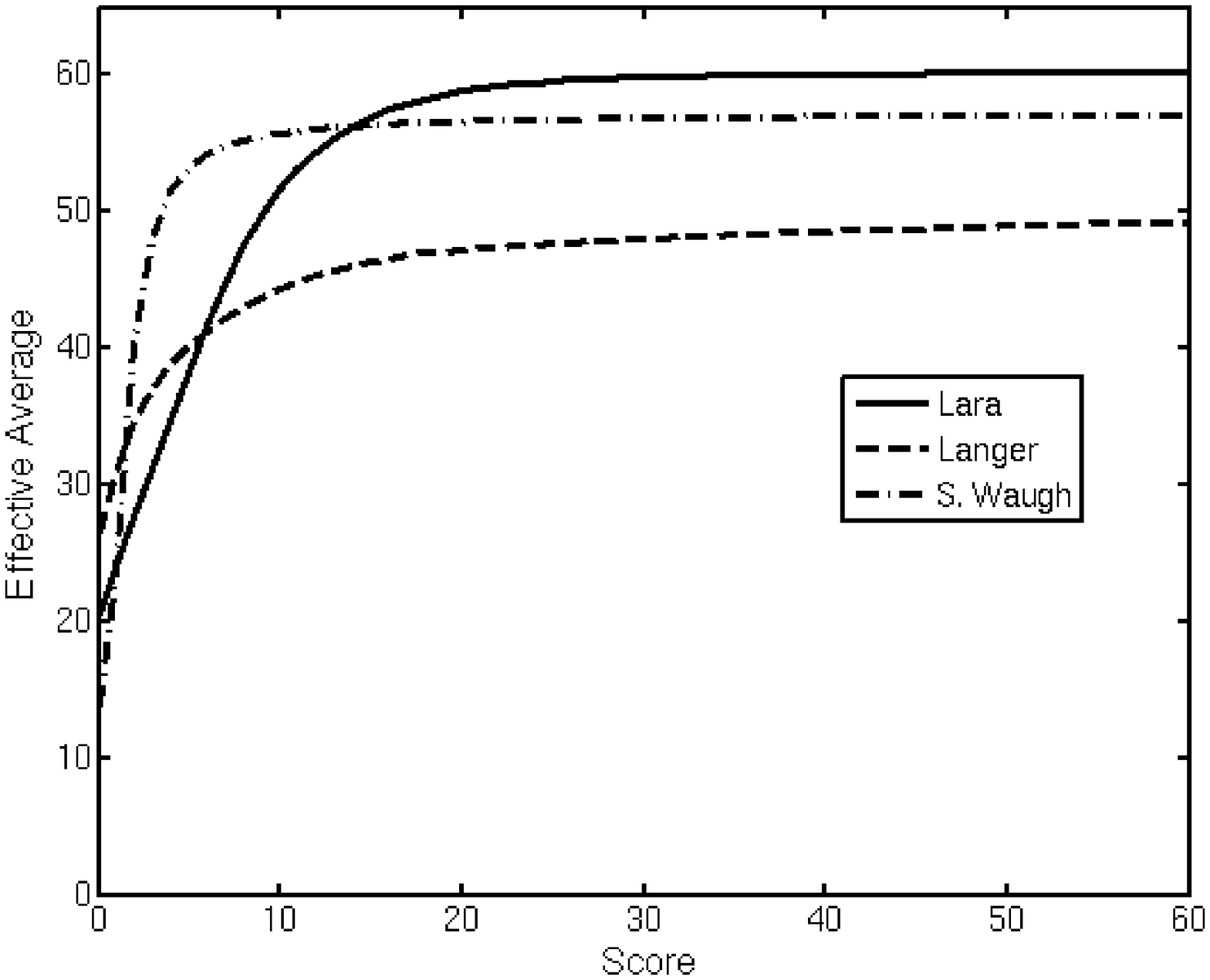}
\end{center}
\caption{Predictive Hazard Functions for Brian Lara, Justin Langer and Steve Waugh. As expected, Justin Langer proves to be less vulnerable than Brian Lara at the beginning of his innings. However, surprisingly, Steve Waugh is more vulnerable than Brian Lara, although he gets his eye in sooner. \label{predictive}}
\end{figure}

This predictive hazard function is plotted (in terms of the effective average) for three players (Brian Lara, Justin Langer and Steve Waugh) in Figure~\ref{predictive}. The latter two are noted for their grit, whilst Brian Lara is considered an aggressive batsman. These different styles may translate to noticable differences in their predictive hazard functions. It is clear from Figure~\ref{predictive} that Justin Langer is more ``robust'' than Brian Lara, as the difference between his abilities when fresh and when set is smaller ($P\left(\left(\frac{\mu_2}{\mu_1}\right)_{Lara} > \left(\frac{\mu_2}{\mu_1}\right)_{Langer}\right) = 0.80$). This is probably a good trait for an opening batsman. On the other hand, Steve Waugh is actually significantly worse when he is new to the crease than Brian Lara ($P(H_{Waugh}(0) > H_{Lara}(0)) = 0.85$). This surprising result shows that popular perceptions are not necessarily accurate, given that many people regard Steve Waugh as the player they would choose to play ``for their life''\footnote{Technically, the correct choice would be to choose a player that minimised the expected loss, where loss is defined as the amount of injury inflicted on the spectator as a function of the batsman's score.}. However, Steve Waugh's predictive hazard function has a transition to its high equilibrium value that is sooner and faster than both Lara and Langer ($P(\tau_{Waugh}<\tau_{Langer} \textnormal{ and } \tau_{Waugh}<\tau_{Lara} \textnormal{ and } L_{Waugh}<L_{Langer} \textnormal{ and } L_{Waugh}<L_{Lara} ) = 0.53$, the prior probability of this is 0.0625). Therefore, perhaps his reputation is upheld, except at the very beginning of an innings.

Note that the general tendency of the effective average to drift upwards as a function of score does not imply that all batsman get better the longer their innings goes on - since once the transition has occurred, our model says that the hazard rate should stay basically constant. Instead, the hazard function of the predictive distribution describes a gradual change in our state of knowledge: the longer a batsman stays in, the more convinced we are that our estimate of their overall ability $\mu_2$ should be higher than our prior estimate. This is why there is an upwards tendency in the predictive ability function for all players, even well after the change-point transition is completed.

\section{Conclusions}
This paper has presented a simple model for the hazard function of a batsman in test or first class cricket. Applying the model to data from several cricketers, we found the expected conclusion: that batsmen are more vulnerable towards the beginning of the innings. However, this analysis now provides a quantitative measurement of this effect, showing how significant it is, and the fact that there is substantial variation in the size of the effect for different cricketers. Surprisingly, we found that Steve Waugh was the second most vulnerable player in the sample at the beginning of an innings - only Shane Warne, a bowler, was was more vulnerable. Even Shaun Pollock is better at the beginning of his innings. This surprising result would have been very hard to anticipate.

From this starting point, there are several possible avenues for further research. One interesting study would involve much larger samples of players so we can identify any trends. For instance, is it true that all-rounders are more robust batsman in general, or are Chris Cairns and Shaun Pollock atypical? Also, it should be possible to create a more rigorous definition of the notion of robustness discussed above. Once this is done, we could characterise the population as a whole, and search for possible correlations between batting average, strike-rate (runs scored per 100 balls faced) and robustness. Modelling the entire cricket population would also allow for a more objective choice for the parameterisation of the hazard function, and the prior distribution over its parameter space. Depending on the results, these kinds of analyses could have implications for selection policy, especially for opening batsmen where consistency is a highly desirable trait.

\bibliographystyle{plainnat}

\newpage

\end{document}